\documentstyle[11pt,aaspp4]{article}
\lefthead{Rand}
\righthead{Spectroscopy of DIG in NGC 5775}

\slugcomment{}

\begin{document}

\title{Ionization, Kinematics, and Extent of the Diffuse Ionized Gas
Halo of NGC 5775}

\author{Richard J. Rand\altaffilmark{1}}
\affil{Univ. of New Mexico, Dept. of Physics and
Astronomy, 800 Yale Blvd, NE, Albuquerque, NM 87131}
\authoremail{rjr@gromit.phys.unm.edu}
\altaffiltext{1}{Visiting Astronomer, National Optical Astronomy Observatories,
Tucson, AZ}

\begin{abstract}

We present key results from deep spectra of the Diffuse Ionized Gas
(DIG) halo of the edge-on galaxy NGC 5775.  [N$\,$II]$\lambda6583$ has
been detected up to $z\approx 13$ kpc above the plane in one of two
vertically oriented long slits -- making this the spiral galaxy with
the greatest spectroscopically detected halo extent in emission.  Key
diagnostic line ratios have been measured up to $z\approx 8$ kpc,
allowing the source of ionization and physical state to be probed.
Ionization by a dilute radiation field from massive stars in the disk
can explain some of the line ratio behavior, but departures from this
picture are clearly indicated, most strongly by the rise of
[O$\,$III]/H$\alpha$ with $z$.  Velocities of the gas in both slits
approach the systemic velocity of the galaxy at several kpc above the
plane.  We interpret this trend as a decrease in rotation velocity
with $z$, with essentially no rotation at heights of several kpc.
Such a trend was observed in the edge-on galaxy NGC 891, but here much
more dramatically.  This falloff is presumably due to the
gravitational potential changing with $z$, but will also depend on the
hydrodynamic nature of the disk-halo cycling of gas and projection
effects.  More detailed modeling of the ionization and kinematics of
this and other edge-ons will be presented in future papers.

\end{abstract}


\section{Introduction}

The Reynolds layer or Warm Ionized Medium of the Milky Way is the
repository of about 90\% of the free electrons in the ISM.  How this
medium is ionized, how its large ($\sim$ 1 kpc; Haffner, Reynolds, \&
Tufte 1999) scale height is maintained, and how far it extends
vertically are some of the main outstanding issues regarding this
diffuse medium.  New progress on these issues is being made with the
Wisconsin H$\alpha$ Mapper (WHAM; Reynolds et al. 1998).  A
complementary approach is to examine this medium in external galaxies,
where it is generally referred to as Diffuse Ionized Gas (DIG).
Narrow-band imaging, spectroscopy, and Fabry-Perot observations have,
in recent years, allowed the brightness, spatial distribution,
emission line ratios, and kinematics of DIG to be studied in many
nearby galaxies (e.g. Golla, Dettmar, \& Domg$\ddot {\rm o}$rgen 1994;
Greenawalt, Walterbos, \& Braun 1997; Hoopes, Walterbos, \& Rand 1999;
Wang, Heckman, \& Lehnert 1997; Rand 1997).

Enhanced ratios of [S$\,$II]$\lambda\lambda$ 6716,6731/H$\alpha$ and
[N$\,$II]$\lambda\lambda$ 6548,6583/H$\alpha$ in the Reynolds layer
and external DIG layers relative to HII regions (e.g. Haffner et
al. 1999; Golla et al. 1994; Ferguson, Wyse, \& Gallagher 1996; Rand
1997) can be explained if the dominant source of ionization is dilute
radiation leaked out of star forming regions from massive stars.
However, recent high-quality data indicate departures from this simple
picture: first, the near constancy with $z$ of [S$\,$II]/[N$\,$II] in
both the Milky Way and external edge-on galaxies (Haffner et al.
1999; Golla et al. 1994; Rand 1998), whereas a rise is expected
(e.g. Domg$\ddot {\rm o}$rgen \& Mathis 1994; Sembach et al. 2000);
second, [S$\,$II]/H$\alpha$ and [N$\,$II]/H$\alpha$ reach values
$\gtrsim 1$ which are difficult for models to reproduce; and third,
[O$\,$III]/H$\alpha$ in the edge-on spiral NGC 891 is found to {\it
rise} with $z$ (Rand 1998), whereas it is expected to fall as oxygen
becomes predominantly singly ionized.

Two explanations for this behavior have been put forth.  While viable,
both are limited by our knowledge of the energizing sources of halos
in general.  First, Reynolds, Haffner, \& Tufte (1999) and Haffner et
al.  (1999) have pointed out for the Reynolds layer that the behavior
of [S$\,$II]/H$\alpha$, [N$\,$II]/H$\alpha$ and [S$\,$II]/[N$\,$II]
could be explained if gas temperature rather than dilution of the
ionizing radiation field (and accompanying changes in the ionization
state of the metals) were the key parameter that changed with $z$,
since the first two ratios are very temperature sensitive while the
third is not.  This idea has yet to be applied to an external galaxy.
The alternative explanation, which has generally been considered for
external galaxies (e.g. Rand 1998; Galarza, Walterbos, \& Braun 1999;
Martin 1997), is a secondary source of ionization, such as shocks
(e.g. Shull \& McKee 1979) or turbulent mixing layers (Slavin, Shull,
\& Begelman 1993) which contribute a fraction of the H$\alpha$
emission which is small but increases with $z$.  The attractive
feature of such ionizing sources is that they can yield a high
[O$\,$III]/H$\alpha$ ratio without necessarily dominating the emission
from the other observable lines.

A second issue well addressed in external galaxies is the vertical
variation of the DIG kinematics.  It has been found in the DIG of NGC
891 that the observed velocities along one vertical slit through the
disk become closer to the galaxy systemic velocity with increasing
$z$, suggesting a fall in the rotation speed (Rand 1997).  HI data
suggest that the effect is widespread in the lower halo (Swaters,
Sancisi, \& van der Hulst 1997).  The rate of falloff will depend on
the galactic potential and the hydrodynamic nature of the disk-halo
gas cycle (Benjamin 2000).

Finally, of great interest is the detectable extent of DIG layers.
DIG in NGC 891 and NGC 5775 has been detected to about $z=5$
kpc (Rand 1997; Hoopes et al.  1999), and $z=6$ kpc (Collins et al.
2000), respectively, while deep, wide-field imaging of NGC 4631
indicates emission up to 16 kpc from the plane (Donahue, Aldering, \&
Stocke 1995).

This paper presents spectra of the edge-on galaxy NGC 5775, which has
been previously imaged in the H$\alpha$ line by Collins et al. (2000).
It is an interacting galaxy (e.g. Irwin 1994) with a high far infrared
luminosity and inferred far infrared surface brightness (Collins et
al. 2000) indicating active star formation.  Its DIG layer is very
bright and extended, as mentioned above, and features some of the most
prominent vertical filamentary structure found above a galactic disk.
Here we focus on several dramatic new results from these spectra.  A
full analysis of these data and spectra of three other edge-ons in
terms of the the two ionization/heating scenarios discussed above will be
presented in a future paper (Paper II) by Collins \& Rand (in preparation).
Another paper (Paper III) will examine the kinematics of the NGC 5775
halo.

\section{Observations}

The spectra were obtained at the KPNO 4-m telescope on 1999 June
10--13.  The slit positions run perpendicular to the plane of the
galaxy, as shown in Figure 1, where they are overlaid on the H$\alpha$
image from Collins et al. (2000).  The central positions of the slits
are given in Table 1.  The slit length is 5', the slit width is 2''
for the galaxy observations, and the spatial scale is 0.69" per pixel.
Slit 1 was chosen to include emission from the most prominent
extraplanar DIG filament, while Slit 2 covers a region of weaker
emission.  The KPC-007 grating was used with the T2KB 2048x2048 CCD,
providing a dispersion of 1.42 $\AA$ per pixel, a resolution of 3.5
$\AA$, and a useful coverage of $4700-6800 \AA$.  Many half-hour
spectra were taken.  Total integration times are given in Table 1.

\placefigure{fig1}

The basic reduction steps are as described in Rand (1998), and here we
only describe processing of the final, stacked, calibrated spectra.
Emission from some of the stronger lines extends nearly to the edge of
the slit, complicating sky subtraction.  In addition, there is a focus
variation of unknown origin, causing the lines towards the ends of the
slit to be significantly broader than those closer to the center,
precluding accurate subtraction of the night sky lines.  While the
general sky background can still be corrected for, sky lines, when
blended with galaxy emission lines, are deconvolved in the line
fitting process.  This blending was mainly a problem for the H$\alpha$
line, which coincided with a blended pair of sky lines.  Towards the
ends of the slit, the combined intensity from these sky lines became
very constant, indicating that there was no measurable H$\alpha$ and
that this intensity value could be used in determining the true
H$\alpha$ intensity elsewhere along the slit.  A much fainter sky line
near [O$\,$III] was corrected for in the same way.  Strong sky lines
also precluded measurements or meaningful upper limits on the [O$\,$I]
$\lambda6300$ and He$\,$I$ \lambda5876$ lines for Slit 1, and the
[N$\,$II] $\lambda5755$ line for both slits.

Line parameters reported here are determined from spectra averaged
along the spatial direction, typically by 10 pixels, or 830 pc at an
assumed distance of 24.8 Mpc (Irwin 1994).  Line properties were
determined with Gaussian fits, using a linear fit to the continuum on
each side of the line.  For uncertainties on intensities and central
wavelengths, the variance of these quantities along each slit for sky
lines of a range of intensities were determined, and an estimate of
the dependence of these variances on sky line intensity was formed.
This relation was then used to determine uncertainties for the
emission lines of interest.  The detection limit for intensities is
about 10$^{-18}$ erg cm$^{-2}$ s$^{-1}$ arcsec$^{-2}$.

\section{Results}

One of the most important results is the tremendous height above the
plane to which emission can be detected, providing unique information
on the ionization state and kinematics of gas at large distances above
the midplane.  [N$\,$II] $\lambda 6583$ is detected to about $z=13$
kpc and $z=7$ kpc on both sides of Slits 1 and 2, respectively.
H$\alpha$ is detected up to about $z=8-9$ kpc in Slit 1 and $z=6-7$
kpc in Slit 2.  [O$\,$III] $\lambda 5007$ is detected to about $z=8$
kpc on the NE side of Slit 1.  The halo of NGC 5775 therefore has the
greatest spectroscopically detected extent in emission of any spiral
galaxy halo.  The vertical emission profiles will be examined in Paper
II; we merely point out for the current purposes that, for data
averaged over 10 spatial pixels, the H$\alpha$ intensity reaches peak
values, in Slits 1 and 2, respectively, of 1.8 and 6.5 $\times
10^{-16}$ erg cm$^{-2}$ s$^{-1}$ arcsec$^{-2}$, falling by $z=5$ kpc
to 7.5 and 3.8 $\times 10^{-18}$ erg cm$^{-2}$ s$^{-1}$ arcsec$^{-2}$
(NE and SW values have been averaged together).

Figure 2 shows the vertical runs of the line ratios [S$\,$II] $\lambda
6716$/H$\alpha$, [N$\,$II] $\lambda 6583$/H$\alpha$,
[S$\,$II]/[N$\,$II] and [O$\,$III] $\lambda 5007$/H$\alpha$.  Because
of the 85$^{\circ}$ (Irwin 1994) inclination of the galaxy, for points
within 10'' of the midplane ($z\lesssim$ 1200 pc in the figures), the
spatial axis reflects in-plane, highly-inclined disk structure rather
than true height above the plane, and line ratio variations are more
due to the line of sight crossing HII regions and areas between them.
This fact explains why line ratio minima are not always at $z=0$ kpc.

\placefigure{fig2}

For the first two ratios, one sees, for the most part, similar
behavior to many DIG halos previously observed, although these
measurements extend to much larger heights: a general rise with $z$,
with disk values of 0.2--0.3 and halo values of [N$\,$II]/H$\alpha$
reaching $>$1 in places.  In general, these ratios rise more slowly
with $z$ than in NGC 891.  The relatively low and constant values on
the NE side of Slit 1 arise from the bright filament, and may simply
reflect the well established correlation between these ratios and
H$\alpha$ surface brightness (Wang, Heckman, \& Lehnert 1998; Rand
1998).  In common with the East side of the halo in the spectra of NGC
891 (Rand 1997, 1998), [N$\,$II]/H$\alpha$ on the NE side of Slit 1
reaches a maximum at $z=6$ and subsequently falls, although the
maximum is reached at only $z=2$ kpc in NGC 891.  There is a
suggestion of similar behavior in both lines on the SW side of Slit 1.
[S$\,$II]/[N$\,$II] is reasonably constant in the halo at a value of
about 0.7 in both slits, but shows more variation than in NGC 891,
reaching extremes of 0.4 and 1.1.

We examine [O$\,$III]/H$\alpha$ rather than [O$\,$III]/H$\beta$ since
the former can be determined to larger $z$.  With the exception of the
NE side of Slit 2, where it remains relatively constant up to $z=4$
kpc, this ratio shows a clear increase with $z$, from values of
0.05--0.4 in the disk to as high as 1.2 in the halo.  Unlike the
afore-mentioned line ratios, Slit 1 does not show a significant local
maximum in this ratio, although the final data point on the SW side
suggests that a maximum may have been reached.  The behavior is
qualitatively similar to NGC 891 (Rand 1998), except that those
measurements extended to only $z=2$ kpc, where a value of
[O$\,$III]/H$\beta=0.8$ was reached.  Assuming optically thin gas at
10,000 K, the peak [O$\,$III]/H$\alpha$ of 1.2 corresponds to
[O$\,$III]/H$\beta=3.7$.

[O$\,$I]/H$\alpha$ rises from around 0.02 in the disk to 0.06 at $z=2$
kpc on the SW side of Slit 2.  He$\,$I $\lambda 5876$ has been
detected in Slit 2, but only for four ten-pixel averages in the disk
can meaningful values be measured.  The mean of the four
He$\,$I/H$\alpha$ ratios is $0.039 \pm 0.003$ (the uncertainty is the
dispersion of the four values).  There is no significant correlation
of the ratio with H$\alpha$ intensity.

These data allow halo kinematics to be probed to unprecedented
heights.  Figure 3 shows velocity centroids of the [N$\,$II] $\lambda
6583$ and [S$\,$II] $\lambda\lambda 6716, 6731$ lines as a function of
slit position for data averaged over ten pixels.  The heliocentric
systemic velocity, as determined from HI data, is $v_{sys} = 1680$ km
s$^{-1}$ (Irwin 1994).  Once again, points at $z\lesssim$ 1.2 kpc
represent disk emission.  The midplane values in both slits indicate
rotation speeds roughly consistent with the rotation curve found by
Irwin (1994).  The falloff towards $v_{sys}$ for $z\lesssim$ 1.2 kpc
is expected for a differentially rotating disk viewed at a not quite
edge-on aspect.

\placefigure{fig3}

Beyond this point, though, the velocities are of the halo gas.  They
continue to move closer to $v_{sys}$ with increasing $z$, coming to
within 20 km s$^{-1}$ of $v_{sys}$ except on the SW side of Slit 1.
The simplest explanation of this behavior is that the rotation speed
falls with $z$, as suggested for NGC 891 (Rand 1997), but here
becoming roughly consistent with {\it no} rotation at the largest
heights.  The exact translation from observed to rotational velocities
is complicated by the unknown distribution of gas along the
line-of-sight.  This issue will be explored in Paper III.  Given the
symmetry on either side of the two slits, it is unlikely that the
kinematics are dominated by the tidal interaction NGC 5775 is
undergoing.  For Slit 1, the steeper falloff on the NE side may
reflect a receding motion of the prominent filament.  In Paper III, we
will attempt to constrain the dependence of $v_{rot}$ with $z$ and
examine whether the falloff is expected for a reasonable mass model,
or whether other hydrodynamical effects (Benjamin 2000) may be
important.

\section{Discussion}

Without carrying out detailed modeling, we can already make
conclusions about the possible source(s) of ionization and physical
conditions in the DIG that may explain the emission line ratios and
their vertical runs.  As in NGC 891 and the Reynolds layer, one can
already conclude that pure photoionization models will not be able to
explain the runs of all the line ratios.  Most problematic are the
increasing values of [O$\,$III]/H$\alpha$ with $z$, in complete
contrast to photoionization models.  Also of concern is the relative
lack of variation in [S$\,$II]/[N$\,$II].  Hence, we are led to
consider the two possible deviations from the pure photoionization
picture discussed in \S 1.  Both may be able to reproduce the trends
in these two line ratios, as well as [S$\,$II]/H$\alpha$,
[N$\,$II]/H$\alpha$, and [O$\,$I]/H$\alpha$.  In Paper II, we will use
data on four edge-ons (NGC 891, NGC 5775, UGC 10288, and NGC 4302) to
ascertain whether one or the other of these models provides a more
adequate description of the data.

Other complications include abundances and the extent to which
depletions (e.g. Howk \& Savage 1999a) are important, as well as
scattering of disk light into the line of sight by extraplanar dust.
Absorption due to dust has been shown to be prevalent in several DIG
halos (Howk \& Savage 1999b).  Modeling of NGC 891 (Ferrara et
al. 1996) suggests that scattered light is a minor contributor to
diffuse halo H$\alpha$ emission (10\% at $z=600$ pc) and declines with
$z$.  Nevertheless, scattered light may be significant in some cases.

Apart from the excitation, the kinematics of gaseous halos and the
extent to which gas can be expelled from the disk are relevant issues
not only to the nature of the disk-halo cycle, but in the
interpretation of QSO absorption line systems.  For instance, Mg$\,$II
absorption is probably due to both halos and disks (Charlton \&
Churchill 1998); consequently possible low-$z$ analogs of absorbers
need to be well characterized at faint levels.

\vspace*{0.5in}

The help of the KPNO staff is greatly appreciated.  We are also
grateful to L. M. Haffner, R. Benjamin, and R. Reynolds for reading a
draft of the manuscript.

\newpage

\begin{table}[htb]
\begin{center}
\caption{Summary of the Observations}
\begin{tabular}{llrc}

Slit & Offset along & Hours of    & Noise\tablenotemark{b} \\
 & Major axis\tablenotemark{a} & Integration & (10$^{-19}$ erg cm$^{-2}$ s$^{-1}$ $\AA^{-1}$ pix$^{-1})$ \\
\tableline
1 & 32'' NW & 6 & 3.1 \\
2 & 20'' SE & 5 & 3.3 \\
\end{tabular}                                      
\end{center}                                       
\tablenotetext{a}{Offsets are from the HI kinematic center determined
by Irwin (1994) at R.A. 14$^{\rm h}$ 53$^{\rm m}$57.6$^{\rm s}$,
Dec. 3$^{\circ}$ 32' 40'' (2000.0).}\tablenotetext{b}{Angular pixel
size is 1.38 arcsec$^2$.  An intensity of $2 \times 10^{-18}$ erg
cm$^{-2}$ s$^{-1}$ arcsec$^{-2}$ corresponds to an Emission Measure of
1 pc cm$^{-6}$}
\end{table}

\newpage

\begin{figure}
\caption{Slit positions overlaid on the H$\alpha$ image of NGC 5775
from Collins et al. (2000).  The cross marks the kinematic center from
Irwin (1994).  Contours show the disk structure.}
\label{fig1}
\end{figure}

\begin{figure}
\caption{Dependence of line ratios on slit position for Slit 1 (filled
circles) and Slit 2 (open circles).  Line ratios plotted are (a)
[S$\,$II] $\lambda 6716$/H$\alpha$, (b) [N$\,$II]
$\lambda$6583/H$\alpha$, (c) [S$\,$II]/[N$\,$II] and (d) [O$\,$III]
$\lambda 5007$/H$\alpha$.  Although the position axis is labeled $z$,
note that points at $z\lesssim$ 1200 pc reflect in-plane structure
more than vertical structure.}
\label{fig2}
\end{figure}

\begin{figure}
\caption{The intensity-weighted mean heliocentric velocities of the
[N$\,$II] $\lambda 6583$ and [S$\,$II] $\lambda\lambda 6716, 6731$
lines as a function of position for (a) Slit 1 and (b) Slit 2.  The
dashed line indicates the systemic velocity.}
\label{fig3}
\end{figure}


\begin{references}

\reference{} Benjamin, R. A., 2000, in Astrophysical Plasmas: Theory,
Codes \& Models, eds. J. Arthur and J. Franco, Rev. Mex. A. \& A., in press.

\reference{} Charlton, J. C., \& Churchill, C. W. 1998, \apj, 499, 181

\reference{} Collins, J. A., Rand, R. J., Duric, N., \& Walterbos, R. A. M. 2000, \apj, in press

\reference{} Domg$\ddot {\rm o}$rgen, H., \& Mathis, J. S. 1994, \apj, 428, 647

\reference{} Donahue, M., Aldering, G., \& Stocke, J. T. 1995, \apj, 450, L45

\reference{} Ferguson, A. M. N., Wyse, R. F. G., \& Gallagher,
J. A. 1996, \aj, 112, 2567

\reference{} Ferrara, A., Bianchi, S., Dettmar, R.-J., \& Giovanardi, C. 1996,
\apj, 467, L69

\reference{} Galarza, V. C., Walterbos, R. A. M., \& Braun, R. 1999, \aj, 118, 2775

\reference{} Golla, G.; Dettmar, R. -J., \& Domg$\ddot {\rm o}$rgen, H. 1994,
\aap, 313, 439

\reference{} Greenawalt, B., Walterbos, R. A. M., \& Braun, R. 1997, \apj,
483, 666

\reference{} Haffner, L. M., Reynolds, R. J., \& Tufte, S. L. 1999, \apj, 523, 223

\reference{} Hoopes, C. G., Walterbos, R. A. M., \& Rand, R. J. 1999, ApJ, 522, 669

\reference{} Howk, J. C., \& Savage, B. D. 1999a, \apj, 517, 746

\reference{} Howk, J. C., \& Savage, B. D. 1999b, \aj, 117, 2077

\reference{} Irwin, J. A. 1994, \apj, 429, 618

\reference{} Martin, C. L. 1997, \apj, 491, 561

\reference{} Rand, R. J. 1997, \apj, 474, 129

\reference{} Rand, R. J. 1998, \apj, 501, 137

\reference{} Reynolds, R. J., Tufte, S. L., Haffner, L. M., Jaehnig, K.,
\& Percival, J. W. 1998, PASA, 15, 14

\reference{} Reynolds, R. J., Haffner, L. M., \& Tufte, S. L. 1999, \apj, 525, L21

\reference{} Sembach, K. R., Howk, J. C., Ryans, R. S., \& Keenan, F. P. 2000, \apj, 528, 310

\reference{} Shull, J. M., \& McKee, C. F. 1979, \apj, 227, 131

\reference{} Slavin, J. D. Shull, J. M. \& Begelman, M. C. 1993, \apj, 407, 83

\reference{} Swaters, R. A., Sancisi, R., \& van der Hulst, J. M. 1997,
\apj, 491, 140

\reference{} Wang, J., Heckman, T. M., \& Lehnert, M. D. 1997, \apj, 491, 114

\reference{} Wang, J., Heckman, T. M., \& Lehnert, M. D. 1998, \apj, 509, 93

\end{references}
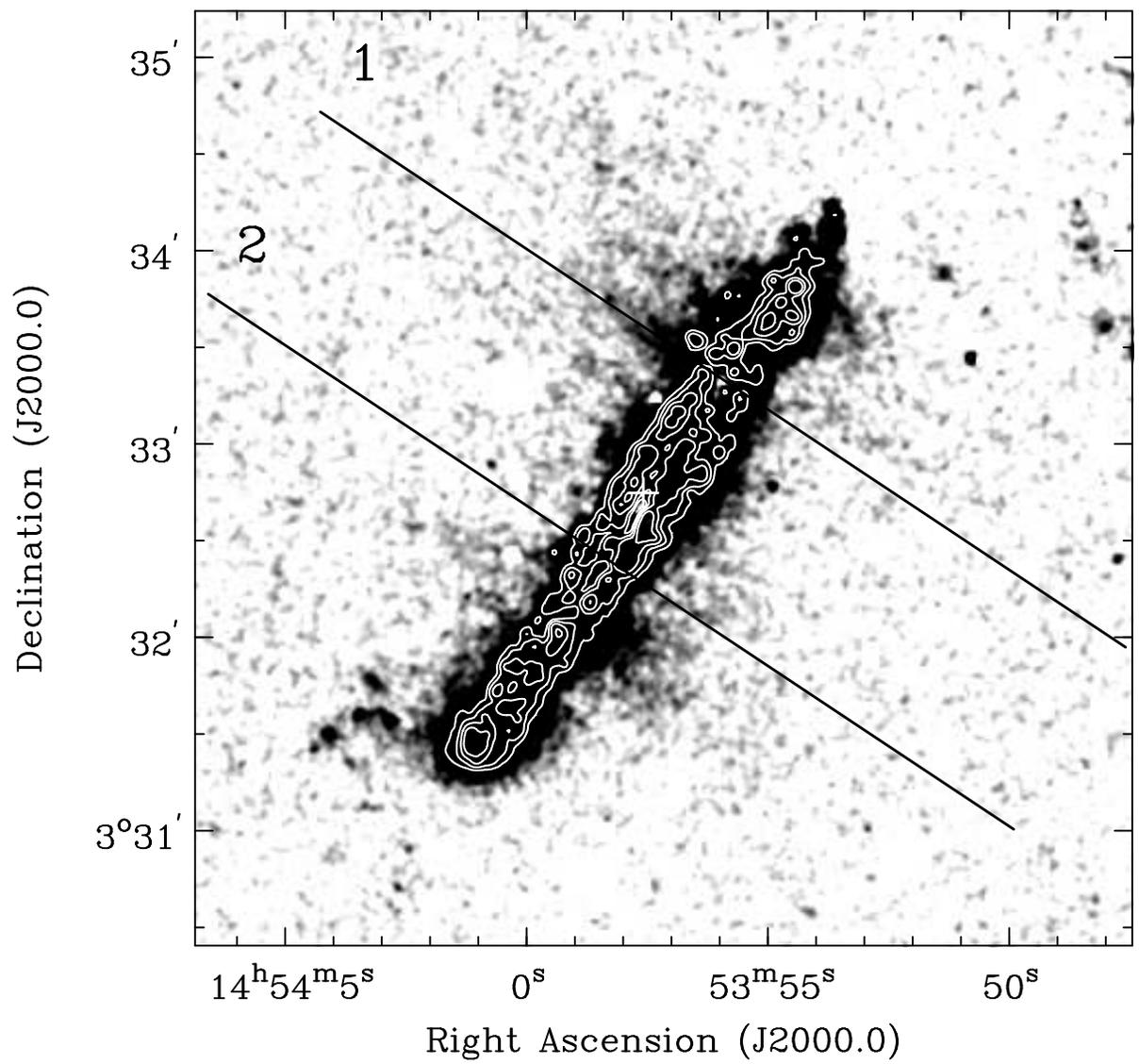
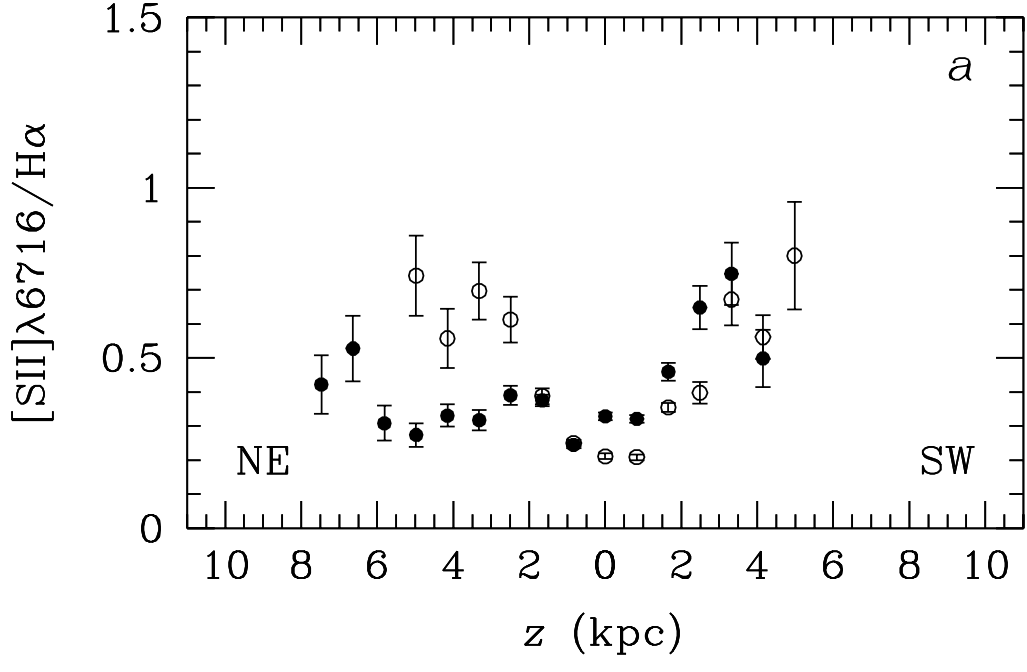
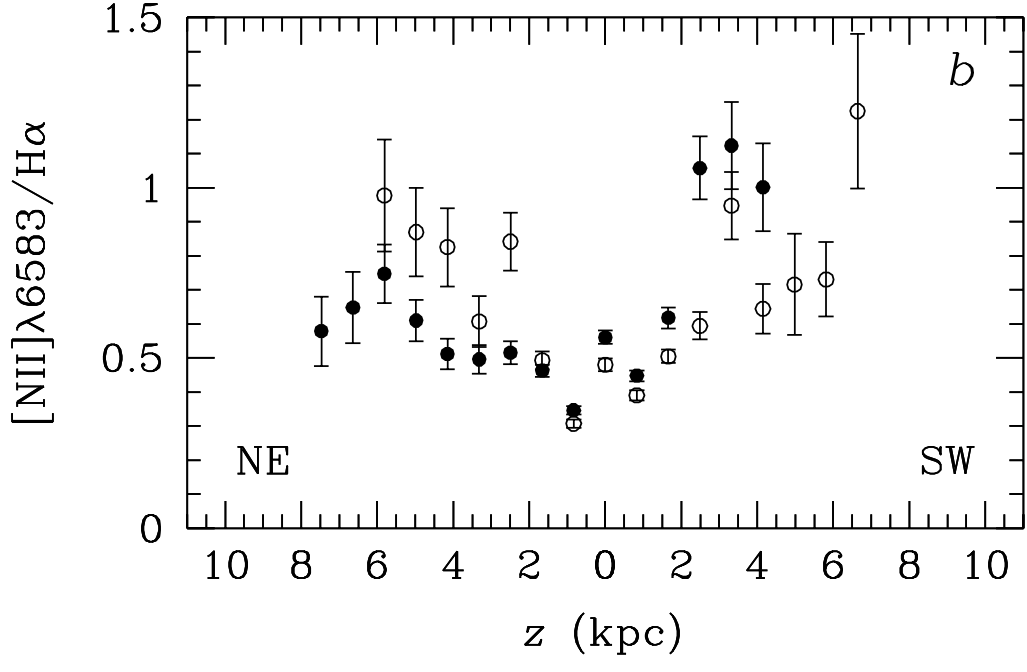
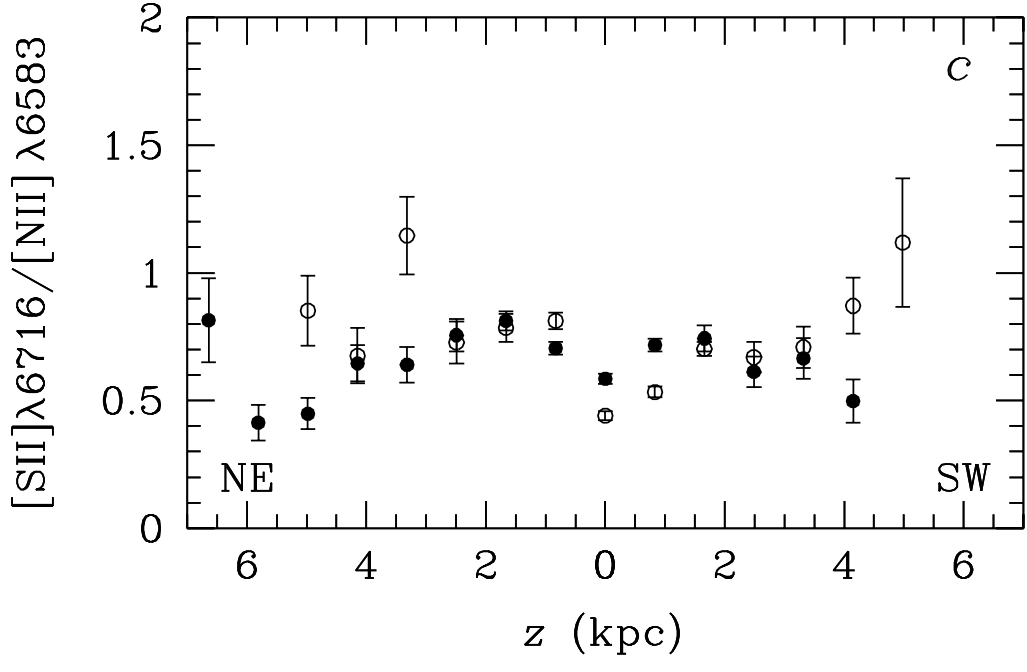
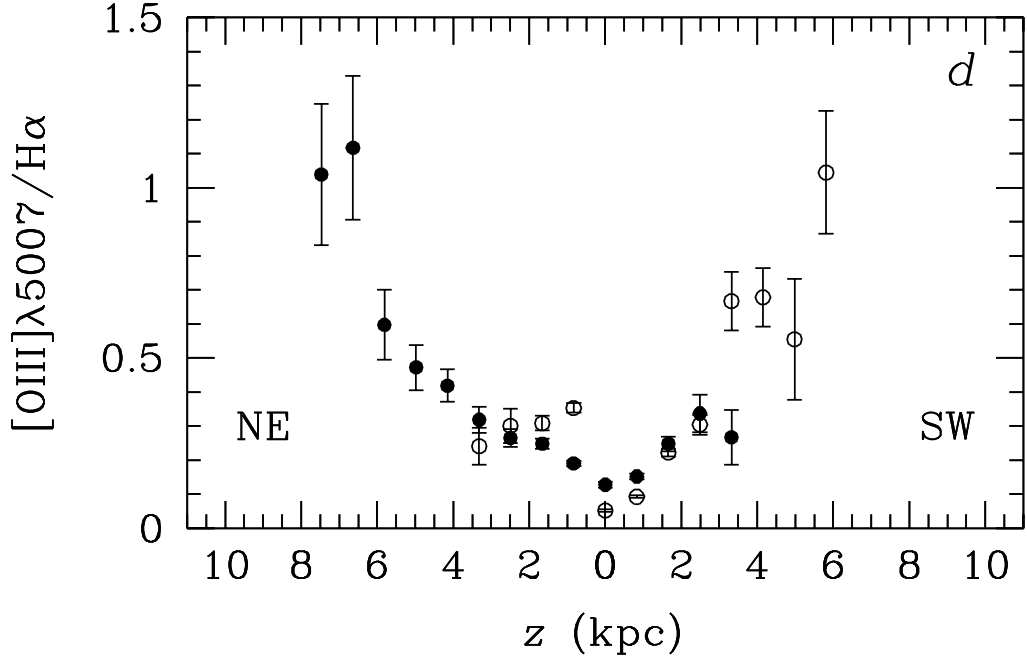
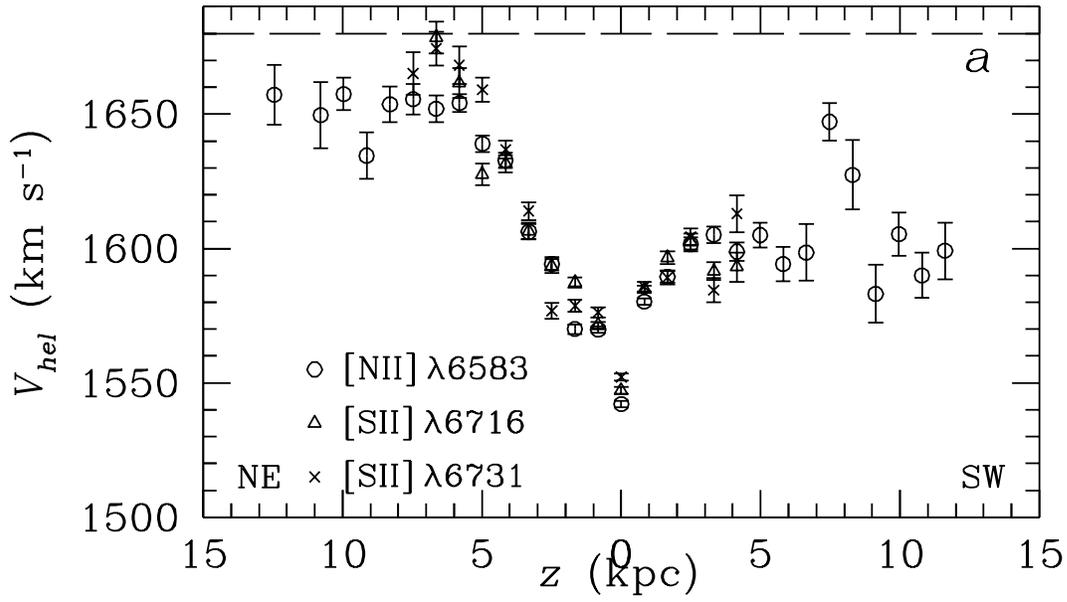
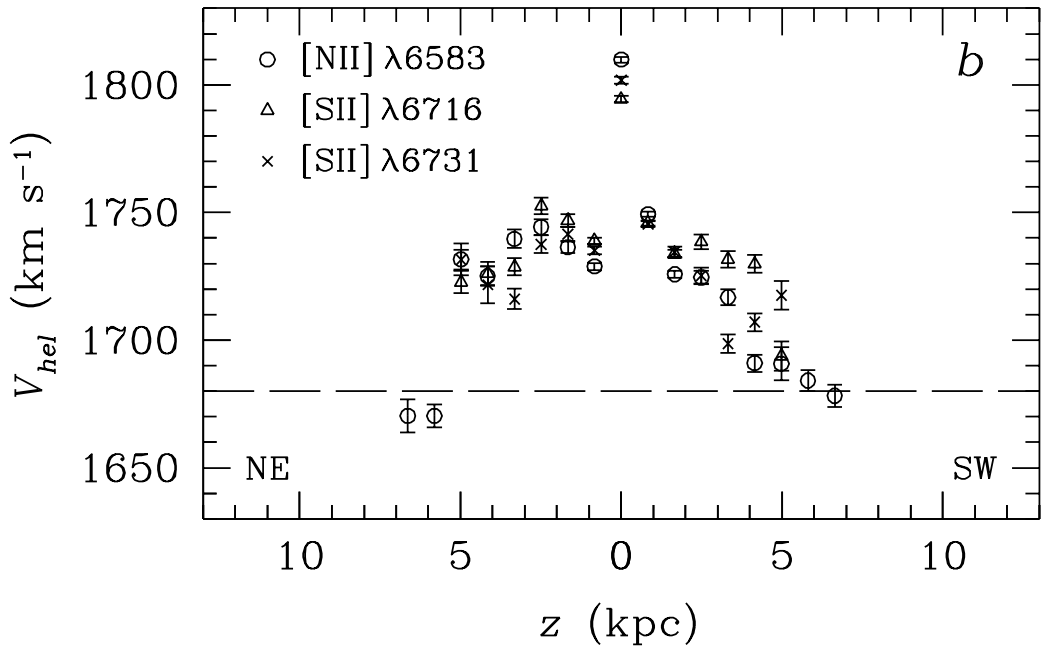
\end{document}